\documentclass[aps,prl,floatfix,twocolumn]{revtex4}
\usepackage{amsmath}
\usepackage{graphicx}
\graphicspath{{figures/}}

\usepackage{units,xspace}
\newcommand{\micron}{\ensuremath{\unit{\mu m}}\xspace}

\renewcommand{\vec}[1]{\ensuremath{{\mathbf #1}}\xspace}
\newcommand{\vecG}{\vec{\Gamma}}
\newcommand{\vecB}{\ensuremath{\vec{B}(\vecG)}}
\newcommand{\hamiltonian}{\ensuremath{\mathcal{H}(\vecG)}}
\newcommand{\order}[1]{\ensuremath{{\mathcal O}(#1)}\xspace}
\newcommand{\Tcon}[1]{\ensuremath{T_\mathrm{con#1}}\xspace}
\newcommand{\TconF}{\ensuremath{T_\mathrm{conF}}\xspace}
\renewcommand{\langle}{\left<}
\renewcommand{\rangle}{\right>}
\newcommand{\avg}[1]{\langle #1 \rangle}
\newcommand{\abs}[1]{\vert#1\vert}

\newcommand{\Th}[1]{\ensuremath{T_h^{(#1)}}}
\newcommand{\Ths}{\Th{s}}

\begin{document}

\title{Anomalous interactions in confined charge-stabilized colloid}

\author{David G. Grier}
\affiliation{Dept.~of Physics and Center for Soft Matter Physics,
New York University,
4 Washington Place, New York, NY 10003}

\author{Yilong Han}

\affiliation{Dept.~of Physics and Astronomy, University of Pennsylvania,
209 South 33rd St., Philadelphia, PA 19104}

\date{\today}

\begin{abstract}
Charge-stabilized colloidal spheres dispersed in weak 1:1 electrolytes
are supposed to repel each other.  
Consequently, experimental evidence for anomalous
long-ranged like-charged attractions induced by geometric confinement
inspired a burst of activity.  This has largely subsided because of
nagging doubts regarding the experiments' reliability and
interpretation.  We describe a new class of thermodynamically
self-consistent colloidal interaction measurements that confirm
the appearance of pairwise attractions among colloidal spheres confined by
one or two bounding walls.  In addition to supporting previous claims for
this as-yet unexplained effect, these measurements also cast new light
on its mechanism.
\end{abstract}


\maketitle


\section{Introduction}

A long-lived controversy was ignited twenty years ago
by the suggestion \cite{sogami83,sogami84}
that similarly charged colloidal spheres need
not repel each other as predicted by Poisson-Boltzmann mean field
theory \cite{derjaguin41,verwey48}, 
but rather might experience a long-ranged attraction for each other
under some circumstances.
Interest in this problem deepened when direct measurements of colloidal
interactions revealed just such attractions in micrometer-scale colloid
in aqueous dispersions at extremely low ionic strength \cite{kepler94}.
Subsequent measurements demonstrated that such anomalous like-charge
attractions are only evident among spheres confined by nearby charged surfaces,
and not otherwise \cite{crocker94,crocker96,crocker96a}.
This observation effectively refuted the originally proposed mechanism
for like-charge colloidal attractions \cite{crocker96a,grier00}, and
other mean-field mechanisms were excluded soon thereafter on
theoretical grounds \cite{neu99,sader99,sader00,trizac00}.

When the search for more sophisticated attraction-generating mechanisms
subsequently failed to reach consensus,
the experimental evidence came under renewed critical scrutiny.
Measurements
of long-ranged attractions performed with optical tweezers
near a single charged wall \cite{larsen97}
were demonstrated to have been sensitive to a previously unsuspected
kinematic coupling mechanism \cite{squires00,behrens01a,behrens01b}.
Suspicion thus was cast on all interaction measurements based on optical
tweezer manipulation in confined geometries \cite{attard01}.
Complementary interaction measurements performed
on colloidal dispersions in equilibrium are immune
to kinematic artifacts
\cite{kepler94,vondermassen94,carbajaltinoco96,bongers98a,behrens01a,brunner02}.
However, they obtain pair potentials by inverting measured pair correlation functions,
a process involving various poorly controlled approximations.
It is conceivable that these methods could misinterpret
oscillatory many-body correlations observed
as attractive or even
oscillatory pair interactions \cite{brunner02}.
Indeed, when particular care was taken to avoid such artifacts in measurements
in a carefully prepared model system,
no sign of anomalous attractions was seen \cite{behrens01a}.
These observations raise a disturbing question:
could the entire case for confinement-induced like-charge attractions be
based on experimental artifacts?

This article describes a new series of equilibrium colloidal interaction
measurements featuring novel tests for thermodynamic self-consistency.
These measurements explicitly address all of the aforementioned sources of
experimental error and yield equilibrium pair potentials (but only where
appropriate!) with quantitative error estimates.
Their results confirm that confinement by one or two nearby glass walls
induces long-range equilibrium attractions 
between nearby pairs of charged spheres.
Confinement-induced attractions appear both among the highly charged polystyrene
sulfate spheres that were the subject of the original round of anomalous observations, 
and also between more weakly charged silica.
Trends observed with variations in confinement and electrolyte concentration
shed new light on the attractions' origin, suggesting a role for
nonmonotonic correlations in the distribution of simple ions near charged surfaces.

\section{The structure of colloidal monolayers}

\begin{figure}[htbp]
  \centering
  \includegraphics[width=0.8\columnwidth]{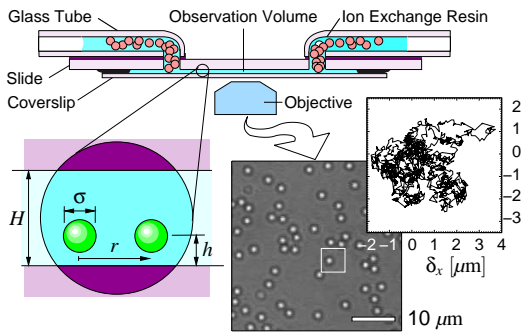}
  \caption{Measuring the structure of colloidal monolayers.}
  \label{fig:schematic}
\end{figure}

Our colloidal interaction measurements follow the general approach pioneered by 
Kepler and Fraden \cite{kepler94} and Vondermassen \emph{et al.} \cite{vondermassen94},
in which digital video microscopy is used to measure the distribution of spheres
in a dispersion at equilibrium.  Figure~\ref{fig:schematic} shows our
implementation schematically.  An aqueous charge-stabilized
dispersion fills a hermetically sealed slit pore between a glass microscope 
slide and a coverslip.  The confined dispersion
is allowed to equilibrate with reservoirs of
mixed-bed ion exchange resin to a base concentration of roughly $1~\unit{\mu M}$.
Controlling the pressure of a buffer gas in these reservoirs also permits the
spacing $H$ between the walls to be adjusted and maintained constant over the
course of an hour \cite{crocker96a}.
Residual contaminant ions are believed to consist of sodium ion leached from the
glass, and carbonate infiltrating from the atmosphere, both of which are monovalent.
The glass surfaces develop large negative charge densities \cite{behrens01b}
that repel negatively charged colloidal spheres and prevent them from sticking
under the influence of van der Waals attraction.  Depending on the resulting
balance of forces on the spheres, the dispersion can be confined
to a monolayer at height $h$ above the lower surface.

Spheres larger than a few hundred nanometers in diameter are readily imaged 
by conventional bright-field microscopy.  A detail from a typical video
micrograph of $\sigma = 1.58~\micron$ diameter silica spheres appears in figure~\ref{fig:schematic}.
The spheres' centers can be tracked with
standard techniques of digital video microscopy \cite{crocker96}, with accuracies
approaching $\Delta x = \Delta y = 30~\unit{nm}$ 
being achieved for these particles \cite{han03a,han04}.
The plot in figure~\ref{fig:schematic} shows the trajectory of a single sphere
over one minute from the region indicated by the box overlaid on the 
micrograph.

The in-plane positions, $\vec{r}_j(t)$, of spheres labeled by $j$ in snapshots
obtained over time $t$ can be compiled into the time-dependent particle density
\begin{equation}
  \label{eq:density}
  \rho(\vec{r},t) = \sum_{j = 1}^{N(t)} \delta(\vec{r} - \vec{r}_j(t)).
\end{equation}
The rest of our results are extracted from $\rho(\vec{r},t)$.

For example, individual trajectories can be analyzed with the Einstein-Smoluchowsky
relation
\begin{equation}
  \label{eq:einstein}
  P(\delta_k|t) = \exp\left( - \frac{(\delta_k - v_k t)^2}{2 D_k t} \right),
\end{equation}
which describes the probability of finding particles displaced by distance
$\delta_k = \avg{\vec{r}_j(t) - \vec{r}_j(0)}_k$ 
along the $k$-th coordinate after time $t$.
Fitting to equation~(\ref{eq:einstein})
yields
the particles' diffusion coefficients $D_k$ and mean
drift velocities $v_k$.
For an equilibrated isotropic system, we expect identical diffusion coefficients
in orthogonal directions and no overall drift.  These conditions are met for
all of the data sets presented below, with maximum drift speeds below
$0.3~\unit{\micron/sec}$ and typical speeds far smaller.

Provided care is taken to account for the finite field of view and the varying
number $N(t)$ of particles within it \cite{bongers98,behrens01a,han03a}, 
$\rho(\vec{r},t)$ can be summarized with
the radial distribution function
\begin{equation}
  \label{eq:gr}
  g(r) = \frac{1}{n^2} \, \avg{\frac{\rho(\vec{r}-\vec{r}^\prime,t) \, \rho(\vec{r}^\prime,t)}{
    A(\vec{r})}},
\end{equation}
where the angle brackets indicate an average over the field of view, over angles, and
over time, and where $n = N/A$ is the areal density of $N = \avg{N(t)}$
particles in area $A$, and $A(\vec{r})$ is the area within the field of view over which
pairs separated by $\vec{r}$ might be found.

\section{Liquid structure inversion}

The Boltzmann formula,
\begin{equation}
  \label{eq:meanforce}
  g(r) = \exp( - \beta w(r) ),
\end{equation}
relates the radial distribution function for an isotropic system in
equilibrium to the
potential of mean force $w(r)$ associated with its structure.
Here, $\beta^{-1} = k_B T$ is the thermal energy scale at absolute
temperature $T$.
The potential of mean force can be identified with the system's
underlying pair potential only in the limit of infinite dilution,
\begin{equation}
  u(r) = \lim_{n \rightarrow 0} w(r).
\end{equation}
At higher densities, simple crowding can induce layering, and thus
oscillatory correlations, even in a system whose pair interactions
are monotonically repulsive.
Interpreting the effective inter-colloid interaction is still more
problematic.  The spheres' dynamics reflect not only their direct
Coulomb repulsions, but also the influence of a sea of atomic scale
simple ions, whose distribution also depends on the spheres' comparatively
enormous charges and excluded volumes.
The effective interaction between two spheres
reflects a thermodynamic average over
the simple ions' degrees of freedom.
This almost certainly will depend on the distribution of other spheres
at higher sphere concentrations.
Under such circumstances, the effective pair potential would not be well
defined.
At lower concentrations, however, the dispersion's free energy
can be described as a superposition of pairwise
interactions.

For all of these reasons, nonmonotonic dependence of 
$\beta w(r) = -\ln g(r)$ on separation $r$
need not signal the onset of attractive interactions.
Particularly in systems with long-ranged repulsive interactions, care
must be taken to correct for many-body correlations.  Unfortunately,
no exact relationship is known between $u(r)$ and $w(r)$ at finite
concentrations, even if the functional form of $u(r)$ is available.
Instead, two strategies
for inverting $g(r)$ have emerged, one involving molecular dynamics
or Monte Carlo simulations to refine trial pair potentials
\cite{kepler94,rajagopalan97}, and another
exploiting results from liquid structure theory 
to correct for many-body correlations \cite{carbajaltinoco96,behrens01a}.
The results from either approach may be identified with the underlying
pair potential thanks to Henderson's uniqueness theorem \cite{duh96}.

We will avail ourselves of the Ornstein-Zernicke 
liquid structure formalism to 
invert $g(r)$ \cite{mcquarrie00}, building upon the pioneering work of
reference~\cite{carbajaltinoco96}.  When applied to the spheres in a colloidal
dispersion, the Ornstein-Zernicke equation describes how effective
interactions among neighboring spheres give rise to structural correlations.
In principle, it describes a hierarchy of $N$-body correlations emerging
from pairwise interactions.  Truncating the hierarchy yields analytically
tractable approximations, whose predictions are increasingly accurate
at lower densities.  Two of these approximations, the hypernetted chain (HNC)
and Percus-Yevick (PY) equations have been found to accurately describe the structure
emerging from computer simulations of systems with long (HNC) and short-range (PY)
interactions.  For two-dimensional systems, these are most conveniently expressed as
\begin{equation}
  \label{eq:liquidstructure1}
  \beta u(r) = \beta w(r) + \left\{ 
    \begin{array}{r}
      n I(r) \qquad \quad {\rm (HNC)}\\
      \ln [1+nI(r)] \qquad \quad {\rm (PY)} 
    \end{array} \right. ,
\end{equation}
where the convolution integral
\begin{equation}
  \label{eq:liquidstructure2}
  I(r) = \int \left[ g(r^\prime) -1 -n I(r) \right]
  \left[ g(\abs{\vec{r}^\prime - \vec{r}})-1  \right] d^2r^\prime
\end{equation}
can be solved iteratively, starting with $I(r) = 0$ \cite{chan77}.
Evaluating $I(r)$ directly rather than with numerical
Fourier transforms 
minimizes the sensitivity of $u(r)$ to noise in $g(r)$.
This implementation has been shown to be both accurate
and effective in previous related studies \cite{behrens01a,han03a,han04}.

\section{Interactions and the DLVO theory}

\begin{figure}[htbp]
  \centering
  \includegraphics[width=.9\columnwidth]{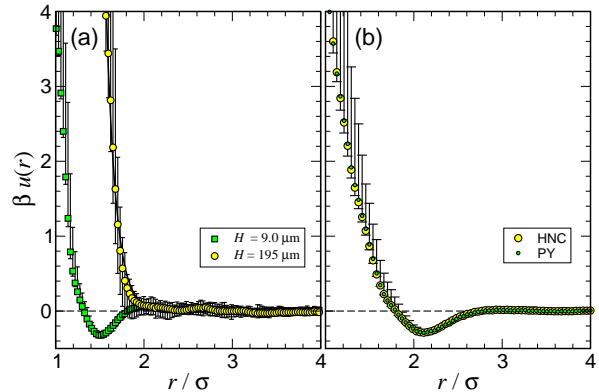}
  \caption{Measured interactions in confined monolayers of (a) silica spheres
    $\sigma = 1.58~\micron$ in diameter and (b) polystyrene spheres 
    $\sigma = 0.652~\micron$
    in diameter.  The silica spheres are sedimented into a monolayer
    at height $h = 0.9~\micron$ above the lower wall.  The two data sets were
    obtained at areal density $n \sigma^2 = 0.0654$ for $H = 9~\micron$
    and $n \sigma^2 = 0.0797$ for $H = 195~\micron$.  The polystyrene spheres,
    by contrast, are confined to the midplane between parallel glass walls
    separated by $H = 1.3~\micron$, with $n \sigma^2 = 0.056$.}
  \label{fig:ur}
\end{figure}

Figure~\ref{fig:ur} shows typical results for pair potentials
obtained from measured
radial distribution functions with both the HNC and PY approximations.
The data plotted as circles in figure~\ref{fig:ur}(a) were
obtained for silica spheres $\sigma = 1.58~\micron$
in diameter in slit pore of heights $H = 195~\micron$ and $H = 9~\micron$.
Silica's density is twice that of water, and these spheres
sediment into a monolayer with their centers at $h = 0.9~\micron$
above the lower glass wall, with
out-of-plane excursions estimated \cite{behrens01a} to be no greater than
$\delta h = 0.1~\micron$.
This system was originally proposed as a model for studying
attractions mediated by a single wall in equilibrium \cite{behrens01a}.
Indeed, the data obtained for a confined monolayer at $H = 9~\micron$
exhibit a strong and long-ranged attraction \cite{han03a}.
The pair potential measured at $H = 195~\micron$, however, is monotonically repulsive
\cite{behrens01a,han03a}.
This observation raises substantial questions regarding the nature of the
more distant wall's influence.

The purely repulsive potential is described very well
by the screened-Coulomb
form predicted by the classic
Derjaguin-Landau-Verwey-Overbeek linearized mean-field
model for colloidal electrostatic interactions \cite{derjaguin41,verwey48}:
\begin{equation}
  \label{eq:dlvo}
  \beta u(r) = Z^2 \lambda_B \, \left( \frac{\exp(\kappa a)}{1 + \kappa a} \right)^2 \,
  \frac{\exp(- \kappa r)}{r}.
\end{equation}
Here, $Z$ is the effective valence of a sphere of radius $a = \sigma/2$,
$\lambda_B = \beta e_0^2 / (4 \pi \epsilon)$
is the Bjerrum length for a medium of dielectric constant $\epsilon$ at temperature
$T$, where $e_0$ is the elementary charge, 
and $\kappa^{-1}$ is the Debye-H\"uckel screening length given by
$\kappa^2 = 4 \pi \lambda_B n_0$ in an electrolyte with a concentration $n_0$ of
monovalent ions.
Fitting to the $H = 200~\micron$
data in figure~\ref{fig:ur}(a) yields a charge number $Z = 6500 \pm 1000$,
in good agreement with predictions of charge renormalization theory \cite{behrens01b},
and screening length $\kappa^{-1} = 180\pm 10~\unit{nm}$ consistent with the 
system's estimated
micromolar ionic strength.
Comparable results are obtained for monolayers at areal densities ranging
from $n \sigma^2 = 0.04$ to $n \sigma^2 = 0.10$, suggesting 
that the result is independent of density, and that the liquid structure inversion
correctly accounts for many-body correlations in this concentration range.
All other results reported here were obtained under comparable conditions.

The observation of DLVO-like repulsions in a weakly confined silica monolayer
is consistent with previous reports on this system \cite{behrens01a}.
It also demonstrates that
our methods do not necessarily yield nonmonotonic potentials
in this range of experimental conditions.
When viewed in this light, the appearance of an attractive minimum
in the pair potential
for the more tightly confined but otherwise identical monolayer at $H = 9~\micron$
seems more credible that it otherwise might \cite{han03a,han04}.
The observation of attractions in silica colloid breaks the monopoly
on anomalous attractions held by the substantially more highly charged polystyrene
sulfate spheres used in previous studies 
\cite{kepler94,crocker94,crocker96a,carbajaltinoco96}.

Such indirect verification does not make the result any less surprising, 
however.
The potential's minimum is roughly $0.3~k_B T$ deep at a center-to-center
separation of $r = 2.4~\micron$.
The interaction's attractive component thus is substantially longer ranged
that the core electrostatic repulsion and measurably influences colloidal
dynamics a distances extending to several screening lengths.
This greatly exceeds the
range of like-charge macromolecular attractions ascribed to polyvalent counterions,
counterion correlations, or fluctuations in the counterion distribution.
Still more puzzling is that a wall separated from the monolayer by nearly
8~\micron can qualitatively transform the spheres' apparent pair potential.

Comparably strong and long-ranged attractions are evident in 
the data plotted in figure~\ref{fig:ur}(b), which were
obtained for polystyrene spheres $\sigma = 0.652~\micron$ in diameter
confined to the midplane between glass walls separated by $H = 1.3~\micron$.
This is consistent with all previous observations of like-charge attractions
in confined polystyrene \cite{kepler94,carbajaltinoco96,crocker96a}, including
those involving optical tweezers \cite{crocker96a}.

As an additional reliability check, results for the polystyrene data
are plotted using both the HNC and PY approximations.  Their quantitative
agreement suggests that the monolayer's areal density is low enough
for the liquid structure formalism to account accurately for many-body
correlations in $g(r)$.  Indeed, there is little difference between
$w(r)$ and $u(r)$ for this data set.
We calculate the difference $\Delta u_L(r)$
between the HNC and PY approximations for each data set
and add it in quadrature to other sources of uncertainty 
to estimate errors in the reported $u(r)$.

By far the largest source of error results from experimental 
uncertainties in $g(r)$.
These, in turn, result from errors in measuring particle position
and from counting statistics.
Assessing the latter turns out to be somewhat subtle and establishes
the lowest practical areal density $n$ at which a reliable measurement
can be made.

The subtlety hinges on the following question: How many snapshots
are required to ascertain whether or not the particles interact at all?
In other words, how many pairs would we expect to see at the
center-to-center separation $r$ in a non-interacting system?
Given a spatial resolution $dr$ for binning particle separations
into the radial distribution function, 
this number is $2 \pi n^2 A r \, dr$.
Typically, the number $N = nA = \order{100}$ of particles in the field of view $A$ is
so small that the expected number of pairs would be unacceptably small.
Combining data from $M$
statistically independent snapshots
reduces the associated error in $g(r)$ to
$\Delta^{(s)} g(r) = g(r) / (2 \pi n^2 A M r \, dr)$.

Errors due to uncertainties in particle location can be calculated
as $\Delta^{(m)} g(r) = 2 \, \partial_r g(r) \, \Delta x$, where
$\Delta x$ is the error in locating a single particle's centroid in each
dimension.
The radial derivative of $g(r)$ can be computed numerically from the
experimental data, which is binned to resolution $dr$
Typically, $\Delta x \ll dr$, so that 
$\Delta^{(m)} g(r) \ll \Delta^{(s)} g(r)$.

Even though the particles' out-of-plane excursions are small, they
also contribute to errors in $g(r)$ through projection errors, especially
near contact.
Out-of-plane fluctuations $\delta h$ make particles
appear to be closer than they actually are.  The error in
apparent particle separation falls off with separation as $\Delta r = (\delta h)^2/r$.
In practice, we combine this contribution in quadrature with the estimated
error due to inaccuracies in particle tracking, $2 \Delta x$, in computing
$\Delta^{(m)}g(r)$.

Combining $\Delta^{(m)}g(r)$ and $\Delta^{(s)} g(r)$ in quadrature establishes
the range of possible values of $g(r)$ for a given sample, restricted only
by the requirement that $g(r) \ge 0$.
We compute trial pair potentials in both the HNC and PY approximations
using both the upper and lower bounds on $g(r)$
as inputs.  The resulting lower and upper estimates on $u(r)$
then are added in quadrature with the systematic error due to differences in
HNC and PY results to obtain estimates for the upper and
lower error bars on $u(r)$.
Typical results appear in figure~\ref{fig:ur}, and establish that the
minima reported in these data are indeed clearly resolved by our methods, even
if the error bounds near contact are substantial.

\section{Thermodynamic self-consistency: Configurational temperature}
\label{sec:tconfig}

Despite the care taken to estimate and eliminate sources of error in these
measurements,
using Eqs.~(\ref{eq:liquidstructure1}) and (\ref{eq:liquidstructure2})
to interpret experimental data might be criticized for its uncontrolled
approximations: Eqs.~(\ref{eq:liquidstructure1}) and (\ref{eq:liquidstructure2})
can converge numerically
to an answer even when applied well beyond their
domain of validity.  Assessing the bounds of this domain can be problematic if
the form of the pair potential is not known \emph{a priori}.
Applying liquid structure theory to experimental data also 
requires the assumption of pairwise additivity.
Nonadditivity, however, would have no obvious signature in the results.
Other unintended processes such as nonequilibrium
hydrodynamic coupling also can yield reasonable-looking results 
that could be mistaken for an equilibrium
pair interaction \cite{squires01}.
Consequently, the appearance of qualitatively new features in any
particular measurement of $u(r)$ could signal a failure in the method.
For this reason, most published accounts
have relied upon comparisons among several related systems to bolster
their conclusions regarding trends in confinement-mediated interactions.
These comparisons are themselves subject to question because the 
ultraclean chemical environments required for these studies are
difficult to alter in a predictable manner.

To address all such concerns, we have introduced \cite{han04}
methods to assess whether or not a trial pair potential describes
a system's interactions in a thermodynamically self-consistent manner.
Our approach is based on the recently introduced notion of a
\emph{configurational temperature}, which has found widespread applications
in simulations \cite{butler98,delhommelle02},
 but has not previously been applied to experimental
data \cite{han04}.

\begin{figure}[htbp]
\includegraphics[width=\columnwidth]{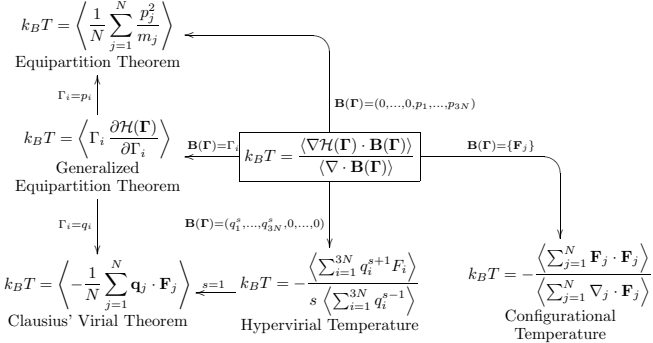}
\caption{Various consequences of the generalized temperature
  definition for selected choices of the arbitrary vector field
  $\vecB$.}
\label{fig:tconfig}
\end{figure}

The temperature of an equilibrium ensemble of particles
is defined conventionally in terms of the particles' mean kinetic energy, 
without regard for their instantaneous positions.
In 1997, Rugh pointed out that the temperature also
can be expressed as
ensemble averages over geometrical
and dynamical quantities \cite{rugh97}.
This notion is expressed more generally \cite{jepps00,rickayzen01} as
\begin{equation}
  k_B T = \frac{\avg{\nabla \hamiltonian \cdot \vecB}}{%
    \langle \nabla \cdot \vecB \rangle},
  \label{eq:general}
\end{equation}
where angle brackets indicate an ensemble average,
$\vecG = \{q_1, \dots, q_{3N}, p_1, \dots, p_{3N}\}$ is the
instantaneous set of $3N$ generalized coordinates $q_j$ and their
conjugate momenta $p_j$ for an $N$-particle system, 
$\hamiltonian = \sum_{j = 1}^{3N} p_j^2 / (2m) + V(\{q_j\})$ is the
Hamiltonian associated with the conservative $N$-particle potential
$V(\{q_j\})$, and $\vecB$ is an \emph{arbitrary} vector field
selected so that both the numerator and denominator of equation~(\ref{eq:general})
are finite and the numerator grows more slowly than $e^N$ in
the thermodynamic limit.
Choosing
$\vecB = \{0, \dots,0, p_1, \dots p_{3N}\}$ yields the
familiar equipartition theorem.
Choosing instead $\vecB = - \nabla V(\{q_i\})$
yields a formally equivalent result,
\begin{equation}
  k_B T_\mathrm{config} = \frac{\avg{\abs{\nabla V}^2}}{%
    \langle \nabla^2 V \rangle},
  \label{eq:tconfig}
\end{equation}
which depends only on the particles' instantaneous configuration,
and not on their momenta.

\begin{figure}[htbp]
  \centering
  \includegraphics[width=0.9\columnwidth]{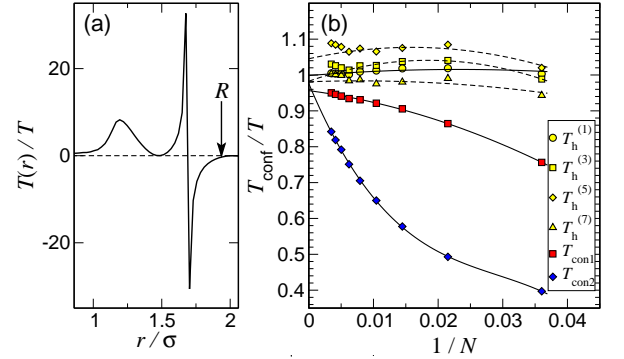}
  \caption{(a) Estimating the range over which interactions affect the configurational temperature
    for the silica data at $H = 9~\micron$.
  (b) Finite-size scaling of several variants of the configurational temperature for the
  silica data at $H = 9~\micron$.  Solid curves are fits to third-order polynomials in $1/N$
  showing extrapolations to the thermodynamic limit.}
  \label{fig:temptr}
\end{figure}

Directly applying equation~(\ref{eq:tconfig}) requires the
full $N$-particle free energy, which is rarely available.
Simplified forms emerge for systems satisfying certain conditions.
For example, if $V(\{q_i\})$ is
the linear superposition of pair potentials, $u(r)$, then
equation~(\ref{eq:tconfig}) reduces to \cite{butler98},
\begin{equation}
  k_B \TconF = - \frac{\avg{\sum_{i=1}^N F_i^2}}{%
    \avg{\sum_{i=1}^N \nabla_i \cdot \vec{F}_i}},
\label{eq:TconF}
\end{equation}
where $\vec{F}_i = -\sum_{j\ne i} \nabla_i u(r_{ij})$ 
is the total force on particle $i$ due to its
interactions with other particles, $\nabla_i$ is the gradient with
respect to the $i$-th particle's position, $\vec{r}_i$,
and $r_{ij} = \abs{\vec{r}_i - \vec{r}_j}$ is the center-to-center
separation between particles $i$ and $j$.
The temperature is reflected in the instantaneous distribution
of forces because objects explore more of 
their potential energy landscape as the temperature increases.

Equation~(\ref{eq:TconF}) may be generalized into a hierarchy of
\emph{hyperconfigurational} temperatures
by choosing $\vecB = \{F_i^s\}$:
\begin{equation}
  k_B \Ths = - \frac{\avg{\sum_{i=1}^N F_i^{s+1}}}{%
    \avg{s \sum_{i=1}^N F_i^{s-1} \, \nabla_i \cdot \vec{F}_i }},
  \label{eq:thyper}
\end{equation}
for $s > 0$.
These higher moments are more sensitive to
the input potential's detailed structure than $\TconF = \Th{1}$.
They also can be applied to three-dimensional systems with long-ranged $1/r$ potentials,
for which \TconF is ill-defined.
Equations (\ref{eq:general}) through (\ref{eq:thyper}) 
apply only in the thermodynamic limit,
with errors of $\order{1/N}$.

For systems with short-ranged potentials, dropping additional terms of $\order{1/N}$ from 
equation~(\ref{eq:general}) yields \cite{jepps00}:
\begin{eqnarray}
  k_B \Tcon{1} & = - \avg{\frac{\sum_{i = 1}^N F_i^2}{%
    \sum_{i = 1}^N \nabla_i \cdot \vec{F}_i}}, \quad {\mathrm and} \label{eq:tcon1}\\
  k_B \Tcon{2} & =  - \avg{\frac{\sum_{i = 1}^N \nabla_i \cdot \vec{F}_i}{%
    \sum_{i = 1}^N F_i^2} }^{-1}, \label{eq:tcon2}
\end{eqnarray}
the second of which was proposed in reference~\cite{han04}.
These definitions' different dependences on sample size $N$ are useful for
comparison with \Ths.

We apply the configurational temperature formalism to our colloidal
monolayers by using the measured particle locations $\rho(\vec{r},t)$
and extracted pair potential $u(r)$ as inputs to the various definitions.
Provided that the conditions for the configurational temperatures'
derivation are met, then all variants
will yield results consistent with
each other and with the (known) temperature $T$ of the heat bath.
In particular, consistent results emerge only if the system is in
local thermodynamic equilibrium, if its interactions
are indeed pairwise additive, and if the measured pair potential $u(r)$
accurately reflects those interactions.

In practice, each snapshot of a monolayer's configuration constitutes
a measurement of its configurational temperature.
Particles near the edge of the field of view, however, may have strongly interacting
neighbors just out of the field of view whose contributions to their net force
would be overlooked.
Including these apparently unbalanced
forces
would grossly distort estimates of the configurational temperature.
To avoid this, we calculate force distributions only for particles
whose relevant neighbors all lie within the field of view.
Such particles lie no closer than the interaction's range $R$ to the edge of the field
of view.  We estimate $R$ from $u(r)$ and $g(r)$ by computing
\begin{equation}
  \label{eq:temprange}
  \frac{T(r)}{T} = 2 \pi \, \beta \, \frac{r}{\sigma} \, g(r) \, 
  \frac{\abs{\nabla u(r)}^2}{\nabla^2 u(r)},
\end{equation}
an example of which is plotted in figure~\ref{fig:temptr}(a).
This function may be interpreted as the contribution
to the configurational temperature due to particles
separated by distance $r$.
Quite clearly, pairs with $r > R$
contribute little if at all to the configurational temperature.

This necessary step further reduces the number $N$ of particles
in the field of view.
This is problematic because all of the temperature definitions involve
approximations of $\order{1/N}$.
Adopting a standard technique from simulation studies, we deliberately
subsample the available data, recalculate the configurational temperature
on the restricted data set, and extrapolate to the large $N$ limit
by fitting the result to a polynomial in $1/N$.
Typical results appear in figure~\ref{fig:temptr}(b).
Even though the different definitions have substantially
different dependences on sample size, they all extrapolate to
the thermodynamic temperature in the thermodynamic limit.

This result turns out to be reassuringly sensitive to details of
the pair potential.
The small residual scatter in the experimental $u(r)$ is greatly
magnified in calculating the configurational temperature, particularly
for the higher-order hyperconfigurational temperatures.  Consequently,
the data in figure~\ref{fig:ur} were fit to a fifth-order polynomial
whose coefficients were used in calculating figure~\ref{fig:temptr}.
Varying the pair potential by as little as one percent in the
region of the core repulsion increases the apparent configurational
temperature by more than ten percent.
Simply truncating the attractive minimum in $u(r)$
to mimic a purely repulsive
potential leads to a fifty percent increase, or an error of $150\unit{^\circ C}$.

The successful collapse of the configurational and hyperconfigurational
temperatures to the thermodynamic temperature constitutes a set
of stringent
internal self-consistency tests for the accuracy of the measured
pair potential and its correct interpretation.
When combined with the considerations from the previous sections,
we can improve the estimated resolution of our pair potential
to roughly $1/20~k_B T$.

The observed confinement-induced attractions therefore should be
considered a real, pairwise additive contribution to the monolayers'
free energy, at least in this range of ionic strength and areal
density.
Attractions of $0.3~k_B T$ are not strong enough to induce
phase separation at such low areal densities, moreover.
This is consistent with the assumption underlying the
liquid structure formalism
that the system is in a single homogeneous phase.

\section{The role of confinement}
\begin{figure}[t!]
  \centering
  \includegraphics[width=0.8\columnwidth]{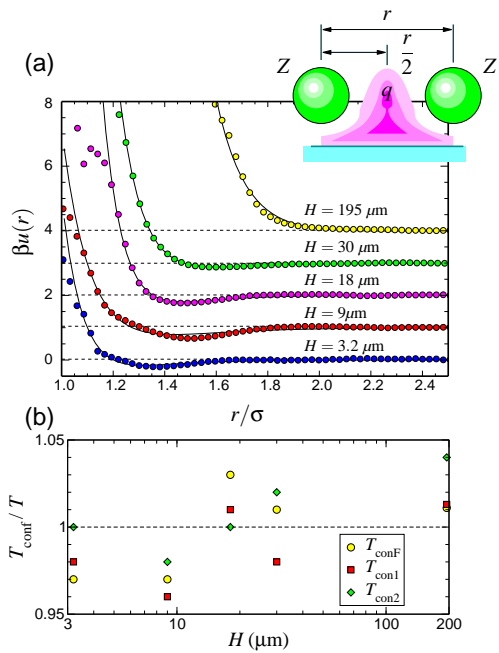}
  \caption{(a) Measured pair potentials for monolayers of silica spheres 1.58~\micron 
    in diameter sedimented to a height $h = 900\pm 100~\unit{nm}$ above a glass surface, 
    for a variety of inter-wall separations, $H$.  The solid curves are
    fits to equation~(\ref{eq:spacecharge}) with parameters tabulated in 
    Table~\ref{tab:spacecharge}.
    The inset schematically represents the space charge model for
    confinement-induced attractions.
    (b) Configurational temperatures for each of the interaction
    measurements.
  }
  \label{fig:spacecharge}
\end{figure}

We next investigate the role of geometric confinement in
inducing like-charge attractions in sedimented silica monolayers.
Figure~\ref{fig:spacecharge} shows data from five different
monolayers of silica spheres ($\sigma =1.58~\micron$) 
in slit pores ranging in depth from 
$H = 195~\micron$ down to $H = 3.2~\micron$.
Figure \ref{fig:spacecharge}(b) shows the associated
configurational temperatures.
For all inter-wall separations, the monolayer is sedimented at
roughly $h = 900~\unit{nm}$, with the only obvious difference being
the inter-wall spacing.
Well-resolved attractive minima are evident for
plate separations as large as $H = 30~\micron$.

This observation contrasts with measurements on more highly
charged polystyrene sulfate spheres, for which anomalous
attractions appear only when the spheres are rigidly confined
to the midplane, at separations no larger
than $H = 4 \sigma$ \cite{crocker96a}.
This difference may be due to the silica spheres' proximity
to the lower wall.  Why then would attractions not be evident
at $H = 200~\micron$ \cite{behrens01a,han03a,han04}?
More to the point, why would a second wall at $H = 20 \sigma$
make a difference?
Trends in figure~\ref{fig:spacecharge}(a) suggest an explanation.

One prominent feature of these data sets is that the apparent range
of the core repulsion moves monotonically to smaller $r$ as the
inter-wall separation decreases.
This differs with the results of optical tweezer measurements on
polystyrene spheres, in which the
depth of the attractive minimum varies with $H$, but not the
range of the repulsion \cite{crocker96a}.
It is tempting to ascribe the trend in our silica data
to a decrease in the effective Debye-H\"uckel
screening length as the ratio of surface area to volume increases
and diffusive contact with the ion exchange reservoirs
diminishes.
If this were the case, however, we would expect the slope of $u(r)$ near contact to
decrease monotonically also.
Instead, there is no discernible trend, presumably because the
base ionic strength varies randomly from run to run.

Referring to the DLVO result in equation~(\ref{eq:dlvo}) for guidance, it would
appear that the spheres' effective charge $Z$ is the only other parameter
that might be free to vary.
Such variation is consistent, at least qualitatively, with predictions
of charge renormalization theory \cite{behrens01b} for silica spheres
near charged silica surfaces.
It would also explain the different behavior of polystyrene spheres whose
more acidic surface groups are not so susceptible to charge regulation
by nearby surfaces \cite{behrens01b}.
It leaves open the question, however, of why an attraction appears at all.

\section{Speculation: Space-charge mediated attractions}

A variety of mechanisms beyond Poisson-Boltzmann mean field theory have been
proposed for confinement-induced attractions among like-charged
colloid.
These include attempts to compute London-like attractions due to
fluctuations in the distribution of simple ions around
the large spheres \cite{gopinathan02} and density functional analysis
of high-order correlations
in the distribution of large and small ions \cite{goulding99a,attard01}.
The few that appear to reproduce experimental observations 
\cite{goulding99,goulding99a} have proved
controversial \cite{trizac99,mateescu01} and none of the more widely accepted calculations
predicts an attraction of the range and
strength observed experimentally, particularly if the simple ions are monovalent.  
Nor have computer simulations yet been able
to address the regime of large charge asymmetry that appears to be necessary
for this effect.
Other approaches, however, may shed light on these anomalous interactions.

The Kirkwood-Poirer formulation of electrolyte structure \cite{kirkwood54},
for example, suggests that the correlations between macroions and simple ions
can become non-monotonic in the strongly coupled regime.
Hastings subsequently pointed out that these correlations in the simple
ion distribution would lead to local violations of electroneutrality
in regions between macroions \cite{hastings78}, and that the resulting
effective interaction between macroions would include an attractive component.
This result parallels the more recent thermodynamically consistent liquid
structure calculation by Carbajal-Tinoco and Gonzalez-Mozuelos \cite{carbajaltinoco02}.

If we hypothesize that the distribution of counterions extending away from
a charged surface also develops regions of space charge when modulated
by nearby spheres, then
the effective inter-sphere interaction should include a term accounting
for sphere-space charge-sphere bridging.
In the absence of a theory for the actual simple ion distribution,
we model the space charge's influence as the screened coulomb interaction
between the spheres' effective charges and a point charge of valence $q$
centered between them:
\begin{equation}
  \label{eq:spacecharge}
  \beta u(r) = Z^2 \lambda_B \, \left( \frac{\exp(\kappa a)}{1 + \kappa a} \right)^2
  \, \frac{\exp(- \kappa r)}{r} 
  - 4 Z q \lambda_B \, \frac{\exp(\kappa a)}{1 + \kappa a}
  \, \frac{\exp\left( - \frac{1}{2} \, \kappa r \right)}{r}.
\end{equation}

Fitting the data in figure~\ref{fig:spacecharge}(a) to this form yields
remarkably good agreement, with fitting parameters tabulated in
Table~\ref{tab:spacecharge}.
The screening lengths in all cases are consistent with the
expected micromolar ionic strengths of our apparatus.
The spheres' effective charge number appears
to decrease systematically with wall separation
in a manner at least qualitatively consistent with charge 
regulation theory \cite{behrens01b}.
Most tellingly, the effective space charge number is
consistent with $q = 10$ at all separations.
If this model is to be taken seriously,
this result suggests that the sedimented silica spheres
are indeed influenced by the nearby wall's counterion
distribution, and that the resulting attraction is evident only
when the core electrostatic repulsion is not too strong.
Reducing the spheres' effective charge exposes the nascent
attraction in this scenario.
For the more highly charged polystyrene spheres, reducing the
wall separation has little effect on the spheres' effective
charge or the screening length, but increases the concentration
of counterions between the spheres.

\begin{table}[htbp]
  \centering
  \begin{tabular}{|r||r|r|r|}
    \hline
    $H$~(\micron) & $Z$ & $\kappa^{-1}$~(\micron) & $q$ \\
    \hline\hline
    195 & $7000 \pm 400$ & $200 \pm 20$ & $3 \pm 3$ \\
    30 & $2500 \pm 150$ & $160 \pm 20$ & $10 \pm 2$ \\
    18 & $2400 \pm 150$ & $140 \pm 20$ & $13 \pm 3$ \\
    9 & $800 \pm 100$ & $150 \pm 25$ & $13 \pm 4$ \\
    3.2 & $800 \pm 100$ & $150 \pm 20$ & $11 \pm 2$ \\
    \hline
  \end{tabular}
  \caption{Interaction parameters obtained from fits to the space charge model.}
  \label{tab:spacecharge}
\end{table}

This simple space-charge model appears to account for the
available observations of like-charge attractions between confined
charge-stabilized spheres.
Its interpretation points toward a correlation-based explanation for
the effect, albeit of an extraordinary range.
The measurements described in the present work should help to eliminate
any remaining concerns regarding the validity and accuracy of 
the larger body of measurements in the literature, and their interpretation.
The thermodynamically self-consistent measurement protocol we introduce
should also find applications in the broader context of experimental soft
matter research.

\section*{Acknowledgments}
This work was supported by the donors of the Petroleum Research
Fund of the American Chemical Society.


\end{document}